# Ferrielectricity in smectic-C* dechiralisation lines lattices

B. Mettout[1], H. Pasco Logbo[2], Y. Gagou[3], H. Vasseur[1] and P.Gisse[1]

[1] *PSC, Université de Picardie, 33 rue saint Leu, Amiens, France*

[2] *Faculté des Sciences et Techniques de Natitingou, USATN - Benin*

[3] *LPMC, Université de Picardie, 33 rue saint Leu, Amiens, France*

Recent experiments probing a new ferroelectric liquid crystal (CLF08) confined in cells with planar alignment have shown dielectric and optic anomalies suggesting the onset of ferrielectric ordering within the surface lattice of dechiralisation lines. We present a phenomenological theory describing the corresponding phase transitions sequence SmA→SmC*→Ferri. Phase diagrams, thermodynamic, dielectric and optic properties are worked out and compared with experiments. The anomalies are related to the predicted tristability of the experimental cells under applied electric field. The order parameters of Landau theory are reinterpreted in terms of line positions, allowing the description of the entrance/exit lines behavior, and yielding the prediction and identification of new *limit* phases within a non-conventional Landau approach.

## I. INTRODUCTION

Bulk and confined properties of chiral SmC* liquid crystals [1,4] have been widely studied since their discovery in the middle of the 70's. They are characterized by their ferroelectric properties at the molecular scale. Although bulk SmC* exhibits an exact helielectric structure [5], in a thin cell with planar geometry the continuous helical symmetry is broken by the presence of the cell walls, and the structure becomes actually anti-ferroelectric at the micron scale. This phenomenon is straightforward at the symmetry point of view and can be more concretely described in terms of dechiralisation lines: The cell walls force the bulk helix to unwind in their vicinity, so that one-dimensional lattices of linear defects appear close to the surfaces in order to protect the helix against the wall unwinding forces. On each side of the sample the distance between two adjacent lines is given by the helix pitch λ. These dechiralisation lines [7,8,9] are positively charged 2π-disclinations of the molecular tilt and polarization fields, which are represented by dots in Fig. 1. Their strong coupling with the applied electric field yields their strong contribution to dielectric properties of the sample [10,11], and in particular to the field-induced unwinding transition [12,13] of the bulk helix. At zero applied electric field the lines are located at the same distance of the walls on both sides of the sample, so that their anti-ferroelectric character becomes clearly apparent: Polarization vectors connecting the negatively charged surfaces to the positive lines alternate periodically along the lattice.

These defects have been observed and described in smectic C samples for more than three decades (though their interpretation is not yet universally accepted [14]), whereas the anti-ferroelectric behavior is evidenced in dielectric measurements by the presence of hysteresis loops of the polarization and permittivity vs. electric field curves exhibiting two typical symmetrical lobes [15,16]. Such a behavior has been recently confirmed [17,18] in a new ferroelectric liquid crystal (CLF08) [19] synthesized by Prof. Dabrowski at Institute of Chemistry, Military University of Technology (Poland), together with a number of anomalous behaviors.

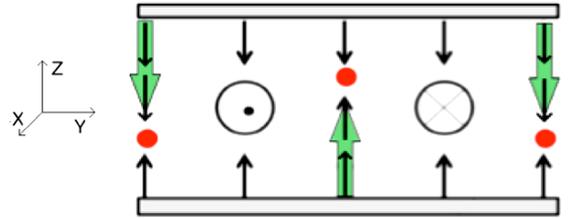

FIG. 1. Scheme of the polarization field in a SmC* cell. Black arrows and circles represent the local polarization. Dots represent cuts of the dechiralisation lines. Close to the surfaces the field is homogeneous. Between the lines the helix is wound. Large arrows show the antiparallel polar vectors defining the globally anti-ferroelectric character of the cell.

Figure 2 presents the real part ε' of the dielectric permittivity versus DC bias voltage in CLF08. The presence of two symmetric lobes gives the hysteresis curve a typical butterfly shape characteristic of anti-ferro. or helielectric materials. The unwinding transition is revealed by a drastic decay of ε' above a critical field. The field hysteresis proves its first-order character.

However, a first anomaly can be seen on this curve: One observes a spectacular difference between the first polarization curve and the cycle: At zero field just after cooling, the permittivity is higher than at the end of the cycle at the same field. Accordingly, there exist two distinct zero-field states, the first one is stabilized when the system is cooled at zero field, the second one is stabilized when the system has been previously submitted to field cycling. The persistency of the hysteresis down to zero field characterizes the second anomaly. It can be seen clearly on the left part of the cycle presented in Fig. 2 as a triangle ending at zero. It becomes more pronounced at lower frequencies.

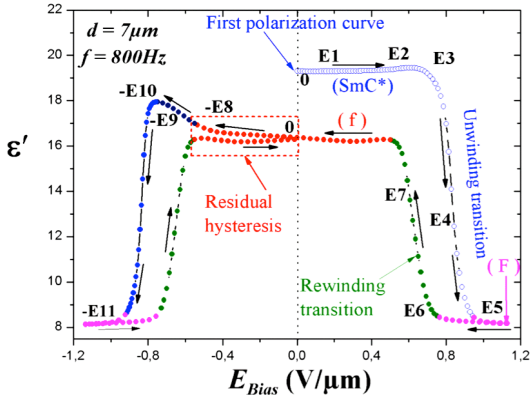

FIG. 2. Hysteresis cycle of the dielectric permittivity ε' versus DC bias voltage $E_{bias}$.

The third anomaly appears as a sharp increase of ε' just before the unwinding transition (-E10 on Fig. 2). The corresponding maximum of ε' becomes stronger at higher frequencies. We interpret this maximum as the trace of a divergence onset characteristic of susceptibility in macroscopically ferroelectric systems, but usually absent in anti-ferroelectrics or helielectrics.

Let us incidentally notice that optic counterparts accompany these dielectric anomalies. The dechiralisation lines are easily observed using polarized light microscope, under crossed polarizer and analyzer. Their behavior during the field cycle, which follows precisely the changes in dielectric properties, will be presented and discussed in section III.

Finally, a fourth anomaly is observed in polarization vs. field P(E) hysteresis loops presented in Fig. 3 for various frequencies. Their shapes change from one-loop curves, typical in macroscopic ferroelectrics [20], to two juxtaposed high-field lobes connected by a central lobe. The two first lobes are associated with the unwinding transition seen in permittivity measurements. On the other hand, the central lobe is associated with the low-field dielectric hysteresis described above in Fig. 2. In addition, we observe that the central lobe width increases on increasing frequencies.

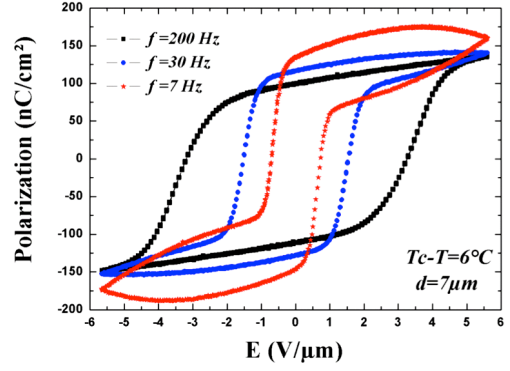

FIG. 3. Polarization versus electric field for different frequencies f.

All these anomalies can be simply interpreted within *the ferrielectric scenario where* macroscopic polarization onsets spontaneously beside the classical SmC* helical polarization field. Thus, the two zero-field states observed in the sample, obtained in the first polarization curve and after cycling, respectively, correspond to distinct phases. The first one is the usual macroscopically non-polarized SmC* phase. The second one (f) is a polarized phase with ferrielectric properties. Moreover, at zero field SmC* is stable and f is metastable. This assumption permits to explain the observed anomalies:

(i) the single low-field loop observed in permittivity and P(E) cycles,

(ii) the divergence (maximum) of the permittivity when the ferrielectric phase approaches its stability limits and,

(iii) the difference of low-field permittivity between the first polarization curve (SmC*) and the cycle (f). In the first polarization curve, the system stays in the stable SmC* phase. On increasing field it undergoes a first-order transition at E3 (Fig. 2) toward an aligned ferroelectric phase with positive polarization ($F^+$) after the SmC* helix unwinding. On decreasing field it undergoes a first-order rewinding transition at E6<E3 toward a domain of the ferrielectric phase with positive polarization ($f^+$), which remains metastable up to zero field. At negative fields (–E10) this state undergoes a transition toward the $F^−$ domain with negative polarization.

During the field cycle: $-E_{max} \to 0 \to +E_{max} \to 0 \to -E_{max}\ldots$, the following periodic phase sequence is predicted:

$$F^- \to f^- \to F^+ \to f^+ \to F^- \ldots ,$$

so that the system never comes again in the SmC* phase.

## II. PHENOMENOLOGICAL THEORY

In order to work out a phenomenological theory accounting for these anomalies, let us analyze the symmetry groups G (zero field) and $G^E$ (under applied field) of the various phases involved in the reordering processes induced by the electric field. Their group/subgroup relationships are presented in Fig. 4(a). Let us first focus on the symmetries of bulk phases (the smectic layers and the helix axis are respectively normal and parallel to Oy, and the field is applied along Oz (Fig. 1). The symbols 1D, 2D and 3D indicate the number of dimensions where continuous translations are present in G and $G^E$. Note that at the micron scale considered in this work translation symmetries normal to smectic layers can be considered as continuous.

- SmA : $G_{A,bulk}$ = $P_{3D}2\infty 2$ (generated by 3D translations, continuous rotations $C_y$ and by twofold rotations axes parallel to the layers). Under electric field, it loses its continuous rotations and becomes:

$$G_{A,bulk}^E = P_{3D}112.$$

- SmC* : $G_{C,bulk}$ = $P_{2D}2\infty_1 2$ (generated by continuous translations along Ox and Oz, by continuous screw axes along Oy and by twofold axes parallel to x and z). Under electric field, the helical axis is broken to discrete translations along Oy, and the group loses its twofold axes, except $C_{2z}$:

$$G_{C,bulk}^E = P_{2D}112.$$

In a cell with planar geometry, the symmetries are broken by the cell walls ($\lambda$ is the helix pitch, that is the distance between two neighboring dechiralisation lines): All translations parallel to Oz and all continuous rotations or screw axes parallel to y are lost.

- SmA : $G_{SmA}$ = $P_{2D}22\infty$ (generated by continuous translations $t_x$, $t_z$ parallel to the cell walls, by the continuous rotations $C_z$ and by the twofold axes $C_{2x}$ and $C_{2y}$). Under electric field:

$$G_{SmA}^E = P_{2D}11\infty.$$

- SmC* (A) : $G_A$ = $P_{1D}22_1 2$ (generated by $t_x$, $C_{2x}$ and $C_{2z}$, the two-fold screw axis $C_{2y}t_{y=l/2}$). Under electric field:

$$G_A^E = P_{1D}112.$$

Since $G_A$ is an antiferroelectric symmetry (the screw axis $2_1$ generates alternating polarization along Oy) we will denote hereafter this phase by the symbol A. It should not be confused with traditional antiferroelectric liquid crystals where the antipolarization arises at the molecular level, while it arises in our system at the micron scale ($\lambda$).

- Ferrielectric : $G_f$ = $G_f^E$ = $P_{1D}112$ (generated by $t_x$, discrete translations $t_{y=\lambda}$ and $C_{2z}$).

- Ferroelectric : $G_F$ = $G_F^E$ = $P_{2D}11\infty$ (generated by $C_z$, and $t_x$, $t_z$).

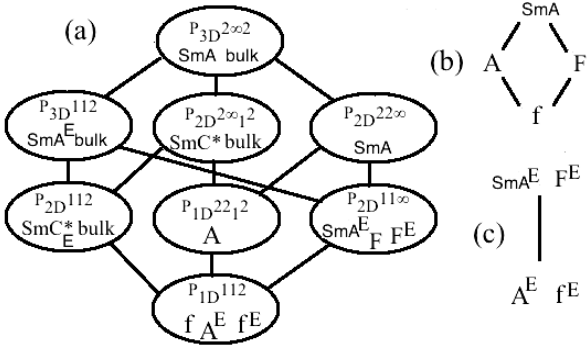

FIG. 4. (a) Group/subgroup relationships between the SmA, SmC* (A), ferro. (F) and ferrielectric (f) phases, in the bulk and in the cell, with and without field. (b) Symmetry breakdown pattern in a cell at zero field. (b) Symmetry breakdown pattern in a cell under electric field.

In the cell the corresponding sequences of symmetry breakdowns (Figs. 4(b,c)) result from the onset of two order parameters [21], $\eta$ and P, representing:

(i) $\eta = \varepsilon e^{i\phi}$ : A transverse polarization (or tilt) wave **p**(y), with: $p_x + ip_z = \eta\, e^{iky}$.

(ii) P : The z-component of the macroscopic polarization.

When an electric field E is applied along Oz, the free energy F depends on E, P and the modulus $\rho$ of the complex wave amplitude $\eta$:

$$F = a\rho^2 + c/2\, \rho^4 + b P^2 + d/2\, P^4 + g/2\, P^6 + K\rho^2 P^2 - E P, \quad (1)$$

where the phenomenological coefficients a,c,b… depend on temperature. F does not depend on the Goldstone angle $\phi$ because it must be invariant under the continuous translation symmetry along Oy of the parent SmA phase. A six-order expansion in P with d<0, g>0 is necessary in order to account for the observed behavior. Indeed, an expansion restricted to the fourth order in P does not allow a metastable ferrielectric phase at zero field.

The equations of state,

$$a\rho + c\rho^3 + K\rho P^2 = 0$$
$$b P + d P^3 + 3g P^5 + 2K\rho^2 P = E, \quad (2)$$

yield at zero field the phase diagram presented in Fig. 5.

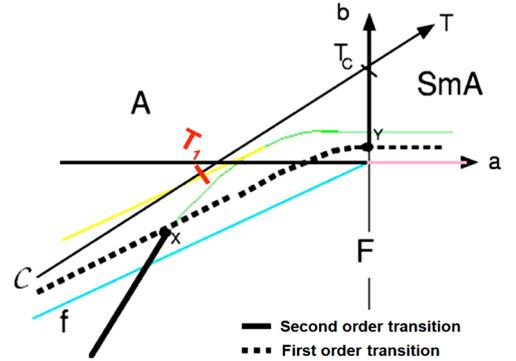

FIG. 5. Zero-field phase diagram of the ferrielectric model. Dotted lines represent first-order transitions. Continuous thick lines represent second-order transitions. Thin lines represent the stability limits of the stable phases: SmA, A (SmC*), ferrielectric f and ferroelectric F. The thin black arrow indicates the thermodynamic path followed by the system when temperature is varied. At low temperature (below $T_1$) the A phase remains stable but coexists with the metastable f phase.

In addition to SmA ($\rho$=P=0), A ($\rho\neq 0$, P=0) and f ($\rho\neq 0$, P$\neq$0), a virtual globally ferroelectric phase F (not to be confused with the locally ferroelectric, but actually antiferroelectric, A phase): $\rho$=0, P$\neq$0 is stabilized. On varying temperature the system follows a thermodynamic path (denoted by C in Fig. 5) that cannot cross the F stability domain. At low temperature the stable A phase coexists with the metastable f phase. Above $T_1$ f becomes unstable and the conventional situation is restored.

Under small constant applied field the previous phase diagram is only slightly modified. The symmetry difference between SmA and F, and the difference between A and f, disappear (see Fig. 4(c)) so that the corresponding first-order transition in Fig. 5 becomes isostructural.

The temperature/field phase diagram along the thermodynamic path C is presented in Fig. 6(a). For fixed temperatures, the corresponding hysteresis P(E) curves are presented in Fig. 6(b).

In fact, when on decreasing field the cycle approaches the field $-E_2$ the system is in the domain $f^+$ of positive polarization of the ferrielectric phase. At $-E_2$ this state becomes unstable and the system can undergo a transition either toward the stable A phase, as is assumed in Figs. 7(a,b), or toward the metastable domain $f^-$ with negative polarization, yielding the cycle ε'(E) shown in Fig. 7(c). Within this scenario only the single low-field loop is predicted, in agreement with observations above 60Hz on P(E) curves (Fig. 3).

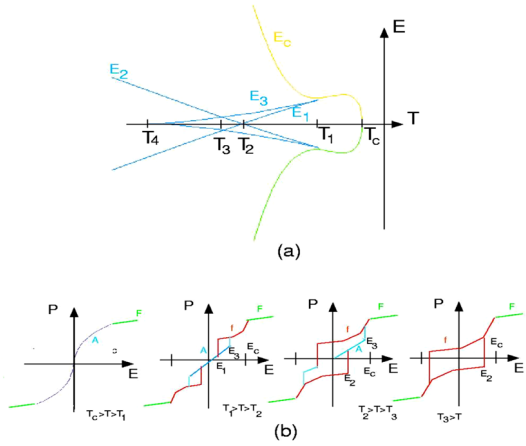

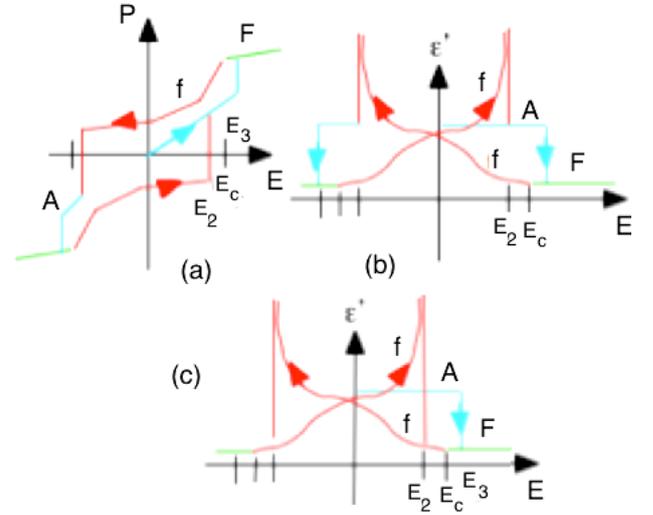

FIG. 6. (a) Temperature/field phase diagram along the thermodynamic path. The thin lines indicate the limits of stability of the various stable or metastable phases. (b) Polarization (P) vs. applied field (E) hysteresis cycles at different fixed temperatures.

FIG. 7. (a) Polarization/field P(E) cycles within the first scenario: Transition $f^+ \to A$ at $-E_2$. (b) Permittivity/field ε'(E) cycle in the first scenario. (c) ε'(E) in the second scenario: Transition $f^+ \to f^-$ at $-E_2$.

Just below $T_c$ only A is stable and no hysteresis can be observed in P(E) curves. Between $T_1$ and $T_2$ a standard double hysteresis is predicted. Between $T_2$ and $T_3$ the system has one stable state, A, and two metastable states, $f^+$ and $f^-$, at zero field. Starting from A on increasing field (first polarization curve) it undergoes a first-order transition to $f^+$, and then a second-order transition to $F^+$. On decreasing field it undergoes a re-winding transition and remains in $f^+$ down to zero field, where it exhibits remnant polarization. Let us notice that the first polarization curve should be distinguished from that observed in a solid ferroelectric, which is formed by a collection of oppositely polarized domains of the spontaneously polar phase. Along the same line, in solids the remnant polarization arises in single domains of the zero-field stable phase while it happens in the metastable f phase in our model.

One has seen that between T2 and T3 the P(E) hysteresis curve has dual character: Single-loop at weak fields and double-loop at high fields, which coincides with the observed behavior (see Fig. 3) except for one point: A two-step transition is predicted from A to F, one at $E_c$ and the other one at $E_3$, whereas observations show a single-step transition.

Accordingly, one has to refine the model in order to get $E_c<E_3$. This can be achieved by letting the thermodynamic path cross, twice, the stability limit of F in Fig. 5. The P(E) hysteresis curve becomes then as in Fig. 7(a). The cycle overlaps a double loop hysteresis at strong fields, characteristic of helielectric states, and a single loop at low fields, characteristic of ferroelectrics. The corresponding static permittivity hysteresis cycle ε'(E) is presented in Fig. 7(b).

The permittivity cycle shown in Fig. 7(c) is also in good agreement with experimental curves shown, for instance, in Fig. 2. Indeed, one observes low-field hysteresis, different behaviors between the cycle and the first polarization curve and one maximum of ε' before the unwinding transition in the cycle. The divergence of ε' predicted in Figs. 7(b,c) occurs at the stability limit of the f phase on increasing field. In fact, since the transition is first order, it can occur before the limit $E_2$ is actually reached, and the divergence should turn into a simple sharp maximum at $E<E_2$ as seen in Fig. 2.

The choice of destination of the transition is a more complex issue than the establishment of phase diagrams, because it also involves the dynamics of the system far from equilibrium. We will not attempt to deal with this difficult problem, and will simply suggest the following heuristic description: When the field is cycled at small speed, the system goes from $f^+$ to the actually stable phase A, leading to a double-loop P(E) curve. At high frequencies, on the contrary, the system transits from $f^+$ to $f^-$ and a single loop is generated. The concomitant decrease in the width of the single-loop on increasing frequency can be explained by the first-order nature of the transition at $E_2$: As stated above, this type of transition can occur actually at any field between E=0 and the *superheating* field $E_2$ corresponding to the metastable limit of $f^+$. The actual value of the transition field depends on defects and dynamics. We therefore assume that at high speed the real transition field goes away from $E_2$ yielding the observed shortening of the single-loop P(E) width (Fig. 3) [17,18].

All the observed qualitative anomalies listed in the introduction are thus well accounted for within our model.

## III. LIMIT PHASES AND DECHIRALISATION LINES

Theoretically, the field-induced unwinding transition in the bulk can be second order. It happens after continuous vanishing of the helix wave vector $k=2\pi/\lambda$. In contrast, in a thin sample we observe that $\lambda$ remains almost constant close to the transition, and the unwinding results from the migration of the dechiralisation lines toward one of the sample walls, followed by their exit out of the liquid crystal. Indeed, they are submitted to Coulomb forces in the direction of the field. Analogously, the rewinding transition results from the entrance and inverse migration of the lines. This process is clearly evidenced in our samples where one can observe under microscope the entrance/exit of each line individually which appears as a dark stripe on the observed surface. The exit begins at one point of the surface cell (Fig. 8) where the line meets a structural defect (focal conic for instance), and one sees then the end point of the line moving along the line until its complete disappearance.

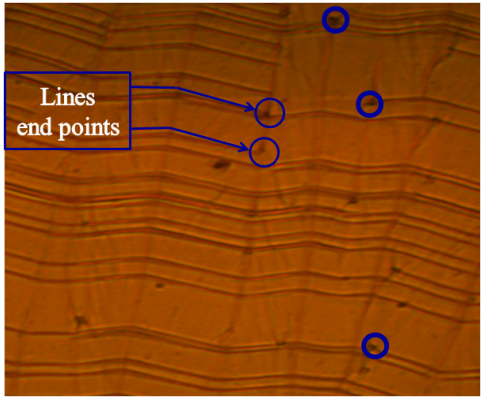

FIG. 8. Sample surface under crossed analyzer/polarizer. The dechiralisation lines appear as horizontal broken parallel dark stripes. The end point of one line (where it leaves the liquid crystal normally to the figure) is indicated.

Observations of lines motion follow precisely the hysteretic dielectric behavior described in section II. The optic observations are schematically summarized in Fig. 9. At zero field just after cooling the lines are symmetrically located with respect to the central plane of the sample and the system is in phase A. No stripe is then observed. Let us denote by H the upper lines and by D the lower lines. On increasing field the lines move toward the field direction and H lines become visible before eventually leaving the sample. Finally, D lines go out the sample, leaving a homogeneous texture at maximum field. On decreasing field the lines come again inside the sample but stay in an unsymmetrical state when the zero field is reached. The system is then in f$^+$ with a positive internal field pushing D lines toward the surface and deforming their structure, making them visible under microscope. On resuming field cycling the exit/entrance line behavior happens without any new stabilization of A. One observes that the texture of the sample is different, on increasing and on decreasing fields, respectively, even at small fields. Thus, optic measurements exhibit anomalies similar to those observed in dielectric experiments: Remnant zero-field polarization and small field hysteresis.

Note that one fact remains unexplained: During the first polarization curve one does not observe any stripe, while the system is in the A phase so that lines should therefore be present. We have no satisfactory arguments to account for this lack of observation, and we have to speculate that a line can only be seen under microscope when it feels an electric field (applied or internal): The field modifies the line structure and brings it closer to the surface, that would make it visible between crossed analyzer and polarizer.

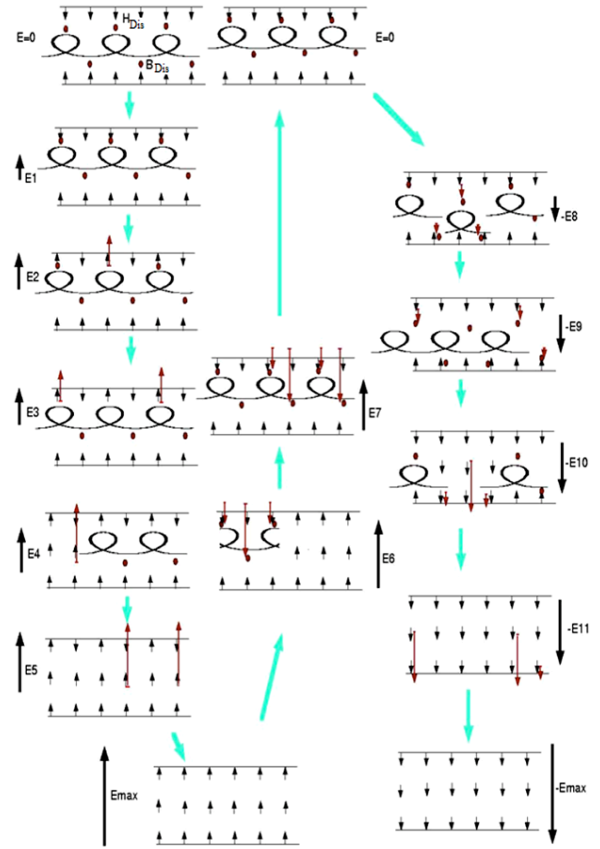

FIG. 9. Schemes of the migration of the dechiralisation lines during field cycling, as observed at $T_C-T=5°$ in a 15μm thick sample from E=0 to $E_{max}$=1V/μm. Schemes in the first row represent the cell during the first polarization curve. The second and third rows represent one half of one cycle. The field values E1, E2 … are defined in Fig. 2. Small arrows represent the polarization field. Large arrows show the displacements of the lines submitted to Coulomb forces. A wrapped line indicates the presence of the helix.

Let us now translate the theory exposed in section II in terms of lines motions. In the previous section, the order parameters ρ and P are formally defined by their transformation properties with respect to the SmA symmetry group, and then related to the molecular polarization field. They can be given different meanings when considering the transitions as resulting from the dechiralisation lines motion. However, this interpretation needs a change in the definition of the parent phase, PP, since in SmA no dechiralisation lines exist. The parent high-symmetry phase is now defined as the locally polarized state where all the lines are located on the sample central plane (z=0). Their distance is then $\lambda/2$ since the corresponding lattice merges both H and D lines (see Fig. 10. (b)). The symmetry group is $P_{1D}222$ characterized by discrete translations $t_{y=\lambda/2}$.

PP has not been experimentally evidenced and might be only virtual in the theoretical phase diagram. However, the possibility of its existence is supported by the following argument: Much above $T_c$ the cell is in the SmA phase, and the polarization (induced by the walls) is non-zero only near the surfaces, beyond which it decreases exponentially. The Lifshitz energy is then not sufficient to overcome the alignment effect of the surfaces and to wind the helix in the polarized surface sheet. The polarization field is normal to the surfaces and remains homogeneous in the directions xy. On approaching $T_c$ the order parameter no longer decreases exponentially from the surfaces. Simultaneously the Lifshitz energy becomes sufficient to modulate its direction along y, which yields the formation of the central lines precursors and breaks the symmetry of the system that transits in the PP phase. Nevertheless, the order parameter remains zero at the sample center (z=0), because it is forbidden by the symmetry group of PP that contains the twofold rotation $C_{2y}$. Above $T_c$, the order parameter begins eventually to take non-zero values at the central plane, which can be achieved only together with actualizing the defect lines on the central plane of the sample and with beginning their migration toward the sample surfaces, yielding the stabilization of the low-symmetry A phase. In this process two second-order phase transitions arise successively SmA→PP→A. It is also possible that they be replaced by a single first-order transition SmA→A, without any stable temperature domain for PP.

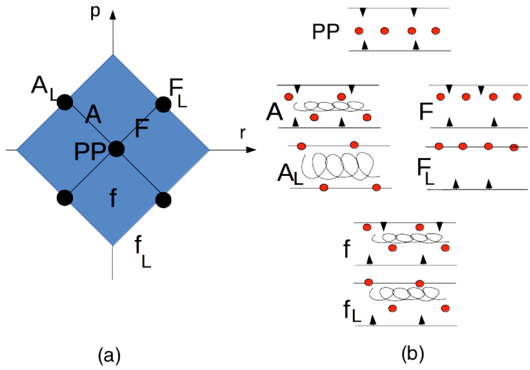

FIG. 10. (a) Order-parameter space. The parent phase PP lies at the center of the square, A and F on the diagonals. The grey area corresponds to the normal ferrielectric f phase. The limit phase $f_L$ lies on the square boundaries. $A_L$ and $F_L$ are represented by dots located on the boundaries. (b) Dechiralisation lines lattices in normal and limit phases. Parts of the sample where the helix is unwound are represented by dark arrows indicating the polarization direction, wrapped continuous thin lines represent the wound helix.

Thus, in the ordered phases A, F and f the lines move away the center. One half of the lines (H) move up while the other half (D) move down, leaving a space to the helix for winding. Denoting by $z_i$ the distance of the i-th line with respect to the center, one can therefore write:

$$z_i = p + (-1)^i r, \quad (3)$$

which defines the new order parameters p and r with the same symmetry than P and ρ. Let us notice that due the discrete translation symmetry of the parent phase along Oy no Goldstone angle ϕ is now present in the description. The lattices of lines are represented in Fig. 10(b) for each stable phase: PP, A, F and f.

The free energy and equations of state given in Eqs. (1,2) remain valid (after replacing P by p and ρ by r). However, an important difference with the previous model results from the fact that the values of p and r are now constrained to stay within the square domain shown in Fig. 10(a). Indeed, when lines reach sample walls, that is, for $z_i = p \pm r = \pm L/2$, then p and r reach the boundaries of this square.

New *limit phases* can thus be stabilized: $A_L$, $F_L$ and $f_L$, located at the square boundaries. The structure of their lines patterns is presented in Fig. 10(b). The existence of such limit phases is an unconventional feature of Landau theory arising when the order parameter belongs to a bounded domain. It is remarkable that, despite the absence of symmetry breakdown between normal and limit phases, the latter's have extended stability domains in the phase diagram, and second-order transitions are permitted with their normal analogs.

Let us note that the fate of the lines after they reach the sample walls is still controversial. In a classical view about defects in condensed matter, they completely disappear and the polarization field remains perfectly aligned everywhere in the sample. In contrast, Pavel *et al* have proposed that a residue persists on the walls, consisting in half integer defects corresponding to π-disclinations.

Two theoretical phase diagrams are presented in Fig. 11 for g=0 and d>0. When $K^2<cd$ only second-order transitions are present, between either group/subgroup related phases, or between normal phases and their limit analogs. When $K^2>cd$ first-order transitions take place between both normal and limit A and F phases.

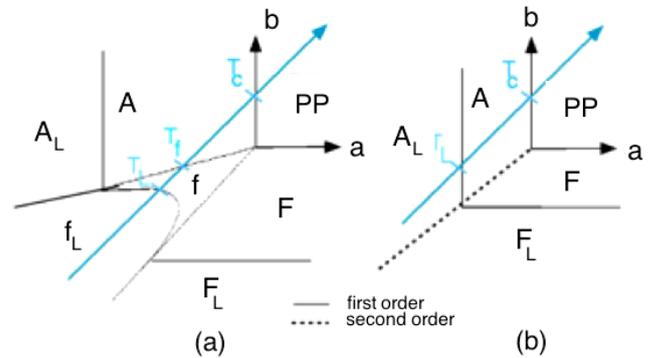

FIG. 11. Zero-field phase diagrams. (a) $K^2<cd$. (b) $K^2>cd$. Arrows indicate possible thermodynamic paths.

Let us now identify the observed states reported in Fig. 9:

(i) During the first polarization process A is stable at zero field. It becomes f between E0 and E1. Between E1 and E2 the exit of H lines transforms the sample into $f_L$. Between E4 and E5 the helix unwinding resulting from the exit of D lines stabilizes $F_L$.

(ii) During the cycling process. Between 0 and E8 the system is ferrielectric $f^+$. Between E8 and E9 it undergoes a transition from $f^+$ to $f^-$. Between E10 and E11 the helix is unwound and all the lines go out the cell, which persists in $F_L$ up to $E_{max}$.

(iii) When the field decreases down to E6, the lines H and D penetrate in the cell simultaneously. Between E6 and E7 the helix is wound and the cell becomes $f^+$ down to zero field.

One sees that during the whole process the system goes through the phases A, $F_L$, f and $f_L$ but never through $A_L$ or F.

Let us illustrate our approach by interpreting the following unexpected behavior observed in 3-μm thick samples at $T_c-T=5°$: During the cycle the sample appears homogeneous at maximum field $E_{max}$. On decreasing field from $E_{max}$, striped domains (i.e. with observable dechiralisation lines) are nucleated and progressively grow. At zero field they occupy half of the total surface. Resuming the cycle for negative fields, these domains decrease and disappear by merging with unstriped regions. It is rather puzzling that these unstriped regions persist during the whole process from $E_{max}$ to $-E_{max}$. They only change progressively their color, which means that the unwinding is not optically visible in these areas.

For explaining this behavior we assume the system is in the part of the zero-field phase diagram (Fig. 5) where A is stable and f metastable. At $E_{max}$ the cell is in $F^+$, the helix is completely unwound and all the dechiralisation lines have disappeared. On decreasing field the system undergoes a first-order transition toward $f^+$ during which growing striped domains coexist with persistent $F^+$ metastable unstriped domains. At lower fields the unstriped domains undergo a second transition toward A which is more stable than f in this part of the phase diagram. Since A is less polarized than f, the internal field is weaker and the lines are no longer visible. Thus, up to zero field unstriped regions (stable A phase) coexist with striped regions (metastable $f^+$ phase). At negative fields the A regions unwind into $F^-$ yielding no optic modifications of the corresponding unstriped regions. Finally only $f^+$ and $F^-$ regions coexist. At large negative fields the ferrielectric phase being less stable than the ferroelectric phase, it gradually declines and leaves the whole sample up to the unstriped unwound F phase.

## IV. CONCLUSION

We interpret the dielectric and optic anomalies observed in the CLF08 SmC* phase below the ordering transition temperature as resulting from the existence of a metastable ferrielectric phase beside the usual A phase (confined SmC*). The tristability (A, $f^+$, $f^-$) of the system under electric field accounts then directly for these anomalies. The helielectric SmC* state is made actually antiferroelectric by the symmetry breaking effects of the sample geometry. The ferrielectric phase results then from the spontaneous onset of a macroscopic polarization in the material. Since such a polarization is forbidden by chirality at molecular scale, it takes necessarily its origin in mesoscopic microstructures, the dechiralisation lines, produced by the unwinding effect of sample surfaces. The spontaneous polarization arising at the corresponding micron scale (λ) is automatically unwound by the surface effects in a few μm-thick samples.

The main effects of the presence of the ferrielectric state, namely low-field hysteresis and remnant polarization, are visible in dielectric measurement as well as under microscope. Indeed, the charged dechiralisation lines strongly interact with the induced internal field and are compelled to follow its hysteretic behavior. However, the converse process is also possible: The existence of spontaneous breakdown of the lines lattice symmetry at zero field stabilizes a structure with ferrielectric group at the micron scale. The dielectric effects result then from the coupling of the polarization field with the lines lattice. We expect this line-based mechanism yields ferrielectricity rather in thin samples where surface effects are dominant. This interpretation is supported by the fact that the observed unwinding transition occurs without any observable change in the lines lattice spacing (i.e. the helix pitch λ), which means the helix/field interactions are not dominant in the unwinding process. Coulomb forces acting on the lines are responsible of their migration across the sample and their exit after the elastic barrier is overcome. We assume the same forces (elastic and electric) between lines and cell walls are responsible of the onset of ferrielectricity in this material.

This theoretical model is consistent and accounts for all the anomalous effects observed in dielectric and optic experiments on CLF08. However, a few points should yet be clarified:

- Why dechiralisation lines are not visible at zero field in the first-polarization curve?

- What are the electric effects of the ions produced by the strong chemical instability of this material? More specifically, can they also explain the observed anomalies?

- Are the observed stripes really traces of the dechiralisation lines?